\documentstyle[preprint,floats,pra,aps]{revtex}
\tightenlines
\begin{document} \draft

\title{\Large \bf Binary Representations of ABCD Matrices}

\author{S. Ba{\c s}kal \footnote{electronic address:
baskal@newton.physics.metu.edu.tr}}

\address{Department of Physics, Middle East Technical University,
06531 Ankara, Turkey}

\author {Y. S. Kim \footnote{electronic address: yskim@physics.umd.edu}}
\address{ Department of Physics, University of Maryland, College Park,
Maryland 20742, U.S.A.}

\maketitle

\begin{abstract}
The $ABCD$ matrix is one of the essential mathematical instruments
in optics.  It is the two-by-two representation of the group $Sp(2)$,
which is applicable to many branches of physics, including squeezed
states of light, special relativity and coupled oscillators.
It is pointed out that the shear representation is oriented to binary
logic which may be friendly to computer applications.  While this is a
future possibility, it is known that para-axial lens optics is based on
the shear representation of the $Sp(2)$ group.  It is pointed out that
the most general form of the $ABCD$ matrix can be written in terms of
six shear matrices, which correspond to lens and translation matrices.
The parameter for each shear matrix is computed
in terms of the three independent parameters of the $ABCD$ matrix.

\end{abstract}

\vspace{5ex}

\narrowtext

\section{Introduction}
In a recent series of papers~\cite{hkn97josa,hkn99}, Han {\it et al.}
studied possible optical devices capable of performing the matrix
operations of the following types:
\begin{equation}\label{shear}
T = \pmatrix{1 &  a \cr 0 & 1} , \qquad
L = \pmatrix{1 &  0 \cr b & 1 } .
\end{equation}
Since these matrices perform shear transformations in a two-dimensional
space~\cite{kiwi90aj}, we shall call them ``shear'' matrices.

However, Han {\it et al.} were interested in computer applications of
these shear matrices because they can convert multiplications into
additions.  Indeed, the $T$ matrix has the property:
\begin{equation}
T_{1} T_{2} = \pmatrix{1 & a_{1} \cr 0 & 1} \pmatrix{1 & a_{2}
      \cr 0 & 1} = \pmatrix{1 &  a_{1} + a_{2} \cr 0 & 1} ,
\end{equation}
and the $L$ matrix has a similar ``slide-rule'' property.  This property
is valid only if we restrict computations to the $T$-type matrices
or to the $L$-type matrices.

What happens if we use both $L$ and $T$ types?  Then it will lead to
a binary logic.  In the present paper, we study this binary property of
the $ABCD$ matrix, which takes the form
\begin{equation}\label{abcd}
G = \pmatrix{ A & B \cr C & D} ,
\end{equation}
where the elements $A, B, C$ and $D$ are real numbers satisfying
$AD - BC = 1$.  Because of this condition, there are three independent
parameters.

We are interested in constructing the most general form of the
$ABCD$ matrix in terms of the two shear matrices given in Eq.(\ref{shear}).
Two-by-two matrices with the above property form the symplectic
group $Sp(2)$.  Indeed, we are quite familiar with the conventional
representation of the two-by-two representation of the $Sp(2)$ group.
This group is like
(isomorphic to) $SU(1,1)$ which is the basic scientific language for
squeezed states of light~\cite{knp91}.  This group is also applicable to
other branches of optics, including polarization optics, interferometers,
layer optics~\cite{monzon99}, and para-axial optics~\cite{kogel66,sudar85}.
The $Sp(2)$ symmetry can be found in many other branches of physics,
including canonical transformations~\cite{kiwi90aj}, special
relativity~\cite{knp91}, Wigner functions~\cite{knp91}, and coupled harmonic
oscillators~\cite{hkn99ajp}.

Even though this group covers a wide spectrum of physics, the mathematical
content of the present paper is minimal because we are dealing only
with three real numbers.  We use group theoretical theorems in order to
manage our calculations in a judicious manner.  Specifically, we use group
theory to represent the most general form of the $ABCD$ matrix in terms of
the shear matrices given in Eq.(\ref{shear}), and to translate the group
theoretical language into a computer friendly binary logic.

With this point in mind, we propose to write the two-by-two $ABCD$
matrices in the form
\begin{equation}\label{chain}
TLTLT~.~.~. .
\end{equation}
Since each matrix in this chain contains
one parameter, there are N parameters for N matrices in the chain.
On the other hand, since both $T$ and $L$ are real unimodular matrices,
the final expression is also real unimodular.  This means that the
expression contains only three independent parameters.

Then we are led the question of whether there is a shortest chain
which can accommodate the most general form of the two-by-two
matrices.  We shall conclude in this paper that six matrices are
needed for the most general form, with three independent parameters.
While we had in mind possible future computer applications of this
binary logic, we are not the first ones to study this problem from
the point of view of ray optics.

Indeed, in 1985, Sudarshan {\it et al.} raised essentially the same
question in connection with para-axial lens optics~\cite{sudar85}.
They observed that the lens and translation matrices are in the form
of matrices given in Eq.(\ref{shear}).  In fact, the notations $L$ and
$T$ for the shear matrices of Eq.(\ref{shear}) are derived from the
words ``lens'' and ``translation'' respectively in para-axial lens
optics.  Sudarshan {\it et al.} conclude that three lenses are needed for
the most general form for the two-by-two matrices for the symplectic
group.  Of course their lens matrices are appropriately separated by
translation matrices.  However, Sudarshan {\it et al.} stated that
the calculation of each lens or translation parameter is ``tedious''
in their paper.

In the present paper, we made this calculation less tedious by using
a decomposition of the $ABCD$ matrix derivable from Bargmann's
paper~\cite{barg47}.  As far as the number of lenses is concerned,
we reach the same conclusion as that of Sudarshan {\it et al.}.
In addition, we complete the calculation of lens parameter for each
lens and the translation parameter for each translation matrix, in
terms of the three independent parameters of the $ABCD$ matrix.

In Sec. \ref{sqsh}, it is noted that the $Sp(2)$ matrices can be
constructed from two different sets of generators.  We call one
of them squeeze representation, and the other shear representation.
In Sec. \ref{decomp}, it is shown that the most general form of the
$Sp(2)$ matrices or $ABCD$ matrices can be decomposed into one symmetric
matrix and one orthogonal matrix.  It is shown that the symmetric
matrix can be decomposed into four shear matrices and the orthogonal
matrix into three.
In Sec. \ref{lensop}, from the traditional point of view, we are
discussing para-axial lens optics.  We shall present a new result in
this well-established subject.
In Sec. \ref{other}, we discuss other areas of optical sciences where
the binary representation of the group $Sp(2)$ may serve useful
purposes.  We discuss also possible extension of the $ABCD$ matrix
to a complex representation, which will enlarge the group $Sp(2)$ to
a larger group.

\section{Squeeze and Shear Representations of the Sp(2)
   Group}\label{sqsh}
Since the $ABCD$ matrix is a representation of the group $Sp(2)$, we
borrow mathematical tools from this group.  This group is generated by
\begin{eqnarray}\label{rot}
&{}& B_{1} = {1\over2}\pmatrix{i & 0 \cr 0 & -i} ,  \qquad
     B_{2} = {1\over2}\pmatrix{0 & i \cr i & 0} ,  \nonumber \\[1.0ex]
&{}& L = {1\over2}\pmatrix{0 & -i \cr i & 0} ,
\end{eqnarray}
when they are applied to a two-dimensional $xy$ space.  The $L$
matrix generates rotations around the origin while $B_{1}$, and
$B_{2}$ generate squeezes along the $xy$ axes and along the
axes rotated by $45^{o}$ respectively.  This aspect of $Sp(2)$ is
well known.  Let us consider a different representation.

The shear matrices of Eq.(\ref{shear}) can be written as
\begin{eqnarray}\label{transf}
&{}& \pmatrix{1 & s \cr 0 & 1} =
             \exp\left(-is X_{1}\right) , \nonumber \\[1.0ex]
&{}& \pmatrix{1 & 0 \cr u & 1} = \exp\left(-iu X_{2}\right) ,
\end{eqnarray}
with
\begin{equation}\label{ex12}
X_{1} =\pmatrix{0 & i \cr 0 & 0}, \qquad
X_{2} =\pmatrix{0 & 0 \cr i & 0} ,
\end{equation}
which serve as the generators.  If we introduce a third matrix
\begin{equation}\label{ex3}
X_{3} = \pmatrix{i & 0 \cr 0 & -i},
\end{equation}
it generates squeeze transformations:
\begin{equation}\label{sqtrans}
\exp\left(-i\eta X_{3}\right) = \pmatrix{e^{\eta} & 0 \cr
0 & e^{-\eta}} .
\end{equation}
The matrices $X_{1}, X_{2}$, and $X_{3}$ form the following closed set
of commutation relations.
\begin{eqnarray}\label{com}
&{}& \left[X_{1}, X_{2}\right] = i X_{3}, \qquad
\left[X_{1}, X_{3}\right] = -2i X_{1}, \nonumber \\[1.0ex]
&{}& \left[X_{2}, X_{3}\right] = 2i X_{2}.
\end{eqnarray}
As we noted in Eq.(\ref{transf}), the matrices $X_{1}$ and $X_{2}$
generate shear transformations~\cite{kiwi90aj,loh93,onc94}.
The matrix $X_{3}$ generate squeeze transformations.  Thus what is the
group generated by one squeeze and two shear transformations?

The generators of Eq.(\ref{ex12})
and Eq.(\ref{ex3}) can be written as
\begin{equation}\label{sp2}
X_{1} = B_{2} - L ,  \quad
X_{2} = B_{2} + L ,  \quad
X_{3} = 2 B_{1} ,
\end{equation}
where $L, B_{1}$ and $B_{2}$ are given in Eq.(\ref{rot}).
The $Sp(2)$ group can now be generated by two seemingly different
sets of generators namely the
squeeze-rotation generators of Eq.(\ref{rot}) and the shear-squeeze
generators of Eq.(\ref{sp2}).  We call the representations generated
by them the ``squeeze'' and ``shear'' representations respectively.
It is quite clear that one representation can be transformed into the
other at the level of generators.  Our experience in the conventional
squeeze representation tells us that an arbitrary $Sp(2)$ matrix
can be decomposed into squeeze and rotation matrices.  Likewise then,
we should be able to decompose the arbitrary matrix into shear and
squeeze matrices.

We are quite familiar with $Sp(2)$ matrices generated by the matrices
given in Eq.(\ref{rot}).  As shown in Appendix A, the most general form
can be written as
\begin{equation}\label{sp22}
G = \pmatrix{\cos\phi & -\sin\phi \cr \sin\phi & \cos\phi}
\pmatrix{ e^{\eta} & 0 \cr 0  & e^{-\eta} }
\pmatrix{\cos\lambda & -\sin\lambda \cr \sin\lambda & \cos\lambda} ,
\end{equation}
where the three free parameters are $\phi, \eta$ and $\lambda$.  The real
numbers $A, B, C$ and $D$ in Eq.(\ref{abcd}) can be written in terms of
these three parameters.  Conversely, the parameters $\phi, \eta$ and
$\lambda$ can be written in terms of $A, B, C$ and $D$ with the condition
that $ AD - BC = 1$.
This matrix is of course written in terms of squeeze and rotation matrices.

Our next question is whether it is possible to write the same matrix
in the shear representation.  In the shear representation, the components
should be in the form of $T$ and $L$ matrices given in Eq.(\ref{shear})
and a squeeze matrix of the form
\begin{equation}\label{sqm}
\pmatrix{ e^{\eta} & 0 \cr 0  & e^{-\eta} } ,
\end{equation}
because they are generated by the matrices given in Eq.(\ref{ex12}) and
Eq.(\ref{ex3}).  But this mathematical problem is not our main concern.
In the present paper, we are interested in whether it is possible to
decompose the $ABCD$ matrix into shear matrices.

\section{Decompositions and Recompositions}\label{decomp}
We are interested in this paper to write the most general form of
the matrix $G$ of Eq.(\ref{abcd}) as a chain of the shear matrices.
Indeed, Sudarshan {\it et al.} attempted this problem in connection
with para-axial lens optics.  Their approach is of course correct.
They concluded however that the complete calculation is ``tedious'' in
their paper.

We propose to complete this well-defined calculation by decomposing
the matrix $G$ into one symmetric matrix and one orthogonal matrix.
For this purpose, let us write the last matrix of Eq.(\ref{sp22}) as
\begin{equation}
\pmatrix{\cos\phi & \sin\phi \cr -\sin\phi & \cos\phi}
\pmatrix{\cos\theta & -\sin\theta \cr \sin\theta & \cos\theta} ,
\end{equation}
with $\lambda = \theta - \phi$.  Instead of $\lambda$, $\theta$ becomes
an independent parameter.

The matrix $G$ can now be written as two matrices, one symmetric and the
other orthogonal:
\begin{equation}\label{gmat}
    G = SR ,
\end{equation}
with
\begin{equation}
R = \pmatrix{\cos\theta & -\sin\theta \cr \sin\theta & \cos\theta} .
\end{equation}
The symmetric matrix $S$ takes the form~\cite{hkn99}
\widetext
\begin{equation}\label{symm}
S = \pmatrix{\cosh\eta + (\sinh\eta)\cos(2\phi) &
     (\sinh\eta) \sin(2\phi) \cr (\sinh\eta) \sin(2\phi) &
     \cosh\eta - (\sinh\eta)\cos(2\phi) } .
\end{equation}
Our procedure is to write $S$ and $R$ separately as shear chains.
Let us consider first the rotation matrix.

In terms of the shears, the rotation matrix $R$ can be written
as~\cite{loh93}:
\begin{equation}\label{rotd1}
R = \pmatrix{1  &  -\tan(\theta/2) \cr 0 & 1}
\pmatrix{1 & 0 \cr \sin\theta & 1}
\pmatrix{1 & -\tan(\theta/2) \cr 0 & 1} .
\end{equation}
This expression is in the form of $TLT$, but it can also be written
in the form of $LTL$.  If we take the transpose and change the
sign of $\theta$, $R$ becomes
\begin{equation}\label{rotd2}
R' = \pmatrix{1  & 0 \cr  \tan(\theta/2)  & 1}
\pmatrix{1 & -\sin\theta \cr 0 & 1}
\pmatrix{1 &  0 \cr \tan(\theta/2) & 1} .
\end{equation}
Both $R$ and $R'$ are the same matrix but are decomposed in  different
ways.

As for the two-parameter symmetric matrix of Eq.(\ref{symm}), we start
with a symmetric $LTLT$ form
\begin{equation}\label{ltlt}
S = \pmatrix{1 & 0 \cr b & 1} \pmatrix{1 & a \cr 0 & 1}
\pmatrix{1 & 0 \cr a & 1} \pmatrix{1 & b \cr 0 & 1},
\end{equation}
which can be combined into one symmetric matrix:
\begin{equation}\label{symm2}
S = \pmatrix{1 + a^{2}  & b(1 + a^{2}) + a
     \cr  b (1 + a^{2}) + a  & 1 + 2ab + b^{2}(1 + a^{2}) } .
\end{equation}
By comparing Eq.(\ref{symm}) and Eq.(\ref{symm2}), we can compute
the parameters $a$ and $b$ in terms of $\eta$ and $\phi$.  The
result is
\widetext
\begin{eqnarray}\label{ab}
 &{}& a = \pm \sqrt{(\cosh\eta - 1) +
              (\sinh\eta) \cos(2\phi)} , \nonumber \\[1.0ex]
 &{}& b = {(\sinh\eta) \sin(2\phi) \mp
\sqrt{(\cosh\eta - 1) + (\sinh\eta) \cos(2\phi)} \over
\cosh\eta + (\sinh\eta) \cos(2\phi)} .
\end{eqnarray}
This matrix can also be written in a $TLTL$ form:
\begin{equation}\label{symm3}
S' = \pmatrix{1 & b' \cr 0 & 1} \pmatrix{1 & 0 \cr a' & 1}
\pmatrix{1 & a' \cr 0 & 1} \pmatrix{1 & 0 \cr b' & 1} .
\end{equation}
Then the parameters $a'$ and $b'$ are
\begin{eqnarray}\label{ab'}
 &{}& a' = \pm \sqrt{(\cosh\eta - 1) -
          (\sinh\eta) \cos(2\phi)} ,  \nonumber \\[1.0ex]
 &{}& b' = {(\sinh\eta) \sin(2\phi) \mp
\sqrt{(\cosh\eta - 1) - (\sinh\eta) \cos(2\phi)} \over
\cosh\eta - (\sinh\eta) \cos(2\phi)} .
\end{eqnarray}
\narrowtext
The difference between the two sets of parameters $ab$ and $a'b'$ is
the sign of the parameter $\eta$.  This sign change means that the squeeze
operation is in the direction perpendicular to the original direction.
In choosing $ab$ or $a'b'$, we will also have to take care of the sign of
the quantity inside the square root to be positive.  If $\cos(2\phi)$ is
sufficiently small, both sets are acceptable.  On the other hand, if
the absolute value of $(\sinh\eta)\cos(2\phi)$ is greater than
$(\cosh\eta - 1)$, only one of the sets, $ab$ or $a'b'$, is valid.

We can now combine the $S$ and $R$ matrices in order to construct
the $ABCD$ matrix.  In so doing, we can reduce the number of matrices
by one
\begin{eqnarray}\label{recomp1}
&{}& SR = \pmatrix{1 & 0 \cr b & 1} \pmatrix{1 & a \cr 0 & 1}
\pmatrix{1 & 0 \cr a & 1}
\pmatrix{1  & b  -\tan(\theta/2) \cr 0 & 1} \nonumber \\[1.0ex]
&{}& \hspace{10ex} \times \pmatrix{1 & 0 \cr \sin\theta & 1}
\pmatrix{1 & -\tan(\theta/2) \cr 0 & 1} .
\end{eqnarray}
We can also combine making the product $S'R'$.  The result is
\begin{eqnarray}\label{recomp2}
&{}& \pmatrix{1 & b' \cr 0 & 1} \pmatrix{1 & 0 \cr a' & 1}
\pmatrix{1 & a' \cr 0 & 1}
\pmatrix{1  & 0 \cr b' + \tan(\theta/2) & 1} \nonumber \\[1.0ex]
&{}& \hspace{10ex} \times \pmatrix{1 & - \sin\theta \cr 0 & 1}
\pmatrix{1 & 0 \cr  \tan(\theta/2) & 1} .
\end{eqnarray}
For the combination $SR$ of Eq.(\ref{recomp1}), two adjoining $T$
matrices were combined into one $T$ matrix.  Similarly, two $L$
matrices were combined into one for the $S'R'$ combination of
Eq.(\ref{recomp2}).

In both cases, there are six matrices, consisting of three $T$
and three $L$ matrices.  This is indeed, the minimum number of shear
matrices needed for the most general form for the $ABCD$ matrix with
three independent parameters.

\section{Para-axial Lens Optics}\label{lensop}
So far, we have been investigating the possibilities of representing
the $ABCD$ matrices in terms of the two shear matrices.  It is an
interesting proposition because this binary representation could
lead to a computer algorithm for computing the $ABCD$ matrix in
optics as well as in other areas of physics.  Indeed, this $ABCD$
matrix has a deep root in ray optics~\cite{kogel66}.

In para-axial lens optics, the lens and translation matrices take
the form
\begin{equation}
L = \pmatrix{1 & 0 \cr -1/f & 1}, \qquad
T = \pmatrix{1 & s \cr 0 & 1} ,
\end{equation}
respectively.  Indeed, in the Introduction, this was what we had
in mind when we defined the shear matrices of $L$ and $T$ types.
These matrices are applicable to the two-dimensional space of
\begin{equation}
\pmatrix{ y \cr m } ,
\end{equation}
where $y$ measures the height of the ray, while $m$ is the slope
of the ray.

The one-lens system consists of a $TLT$ chain.  The two-lens system
can be written as $TLTLT$.  If we add more lenses, the chain becomes
longer.  However, the net result is one $ABCD$ matrix with three
independent parameters.  In Sec. \ref{decomp}, we asked the question of
how many $L$ and $T$ matrices are needed to represent the most general
form of the $ABCD$ matrix.  Our conclusion was that six matrices, with
three lens matrices, are needed.  The chain can be either $LTLTLT$ or
$TLTLTL$.  In either case, three lenses are required.  This conclusion
was obtained earlier by Sudarshan {\it et al.} in 1985~\cite{sudar85}.
In this paper, using the decomposition technique derived from
the Bargman decomposition, we were able to compute the parameter of
each shear matrix in terms of the three parameters of the $ABCD$
matrix.

In para-axial optics, we often encounter special forms of the
$ABCD$ matrix.  For instance, the matrix of the form of Eq.(\ref{sqm})
is for pure magnification~\cite{gerrard75}.
This is a special case of the decomposition given for $S$ and $S'$ in
Eq.(\ref{symm2}) and Eq.(\ref{symm3}) respectively, with $\phi = 0$.
However, if $\eta$ is positive, the set $a'b'$ is not acceptable because
the quantity in the square root in Eq.(\ref{ab'}) becomes negative.
For the $ab$ set,
\begin{equation}
a = \pm \left(e^{\eta} - 1 \right)^{1/2} , \qquad
b = \mp e^{-\eta} \left(e^{\eta} - 1 \right)^{1/2} .
\end{equation}
The decomposition of the $LTLT$ type is given in Eq.(\ref{ltlt}).

We often encounter the triangular matrices of the form~\cite{simon00}
\begin{equation}\label{trian}
\pmatrix{A & B \cr 0 & D}  \quad \mbox{or} \quad
\pmatrix{A & 0 \cr C & D}  .
\end{equation}
However, from the condition that their determinant be one, these
matrices take the form
\begin{equation}
\pmatrix{e^{\eta} & B \cr 0 & e^{-\eta}}   \quad \mbox{or} \quad
\pmatrix{e^{\eta} & 0 \cr C & e^{-\eta}} .
\end{equation}
The first and second matrices are used for focal and telescope
conditions respectively.  We call them the matrices of $B$ and
$C$ types respectively.  The question then is how many shear matrices
are needed to represent the most general form of these matrices.
The triangular matrix of Eq.(\ref{trian}) is discussed frequently in
the literature~\cite{gerrard75,simon00}.  In the present paper, we are
interested in using only shear matrices as elements of decomposition.

Let us consider the $B$ type.  It can be constructed either in
the form
\begin{equation}
\pmatrix{e^{\eta} & 0 \cr 0 & e^{-\eta}}
  \pmatrix{1 & e^{-\eta} B \cr 0 & 1}
\end{equation}
or
\begin{equation}
\pmatrix{1 & e^{\eta} B \cr 0 & 1}
\pmatrix{e^{\eta} & 0 \cr 0 & e^{-\eta}}.
\end{equation}
The number of matrices in the chain can be either four or five.
We can reach a similar conclusion for the matrix of the $C$ type.

\section{Other Areas of Optical Sciences}\label{other}
We write the $ABCD$ matrix for the ray transfer matrix~\cite{gerrard75}.
There are many ray transfers in optics other than para-axial lens
optics.  For instance, a laser resonator with spherical mirrors is
exactly like para-axial lens optics if the radius of the mirror is
sufficiently large~\cite{kahn65}.

If wave fronts with phase is taken into account, or for Gaussian beams,
the elements of the $ABCD$ matrix becomes complex~\cite{kogel65,naka98}.
In this case, the matrix operation can sometimes be written as
\begin{equation}\label{bilin}
w' = { A w + B \over C w + D } ,
\end{equation}
where $w$ is a complex number with two real parameters. This is
precisely the bilinear representation of the six-parameter Lorentz
group~\cite{barg47}.
This bilinear representation was discussed in
detail for polarization optics by Han {\it et al}.~\cite{hkn96pla}.
This form of representation is useful also in laser mode-locking and
optical pulse transmission~\cite{naka98}.

The bilinear form of Eq.(\ref{bilin}) is equivalent to the matrix
transformation~\cite{hkn96pla}
\begin{equation}
\pmatrix{v_{1}' \cr v_{2} '} = \pmatrix{ A & B \cr C & D}
\pmatrix{v_{1} \cr v_{2}},
\end{equation}
with
\begin{equation}
w = {v_{2} \over v_{1}}
\end{equation}
This bilinear representation deals only with the ratio of the second
component to the first in the column vector to which $ABCD$ matrix
is applicable.  In polarization optics, for instance, $v_{1}$ and
$v_{2}$ correspond to the two orthogonal elements of polarization.

Indeed, this six-parameter group can accommodate a wide spectrum of
optics and other sciences.
Recently, the two-by-two Jones matrix and four-by-four Mueller matrix
have been shown to be two-by-two and four-by-four representations of the
Lorentz group~\cite{hkn97josa}.  Also recently, Monz\'on and S\'anchez
showed that multilayer optics could serve as an analog computer for
special relativity~\cite{monzon99}.  More recently, two-beam
interferometers can also be formulated in terms of the Lorentz
group~\cite{hkn00}.

\section*{Concluding Remarks}
The Lorentz group was introduced to physics as a mathematical device
to deal with Lorentz transformations in special relativity.  However,
this group is becoming the major language in optical sciences.  With
the appearance of squeezed states as two-photon coherent
states~\cite{yuen76}, the Lorentz group was recognized as the
theoretical backbone of coherent states as well as generalized coherent
states~\cite{knp91}.

In their recent paper~\cite{hkn99}, Han {\it et al}. studied in detail
possible optical devices which produce the shear matrices of
Eq.(\ref{shear}).  This effect is due to the mathematical identity
called ``Iwasawa decomposition''~\cite{iwa49,simon98}, and this
mathematical technique is relatively new in optics.  The shear matrices
of Eq.(\ref{shear}) are products of Iwasawa decompositions.  Since
we are using those matrices to produce the most general form of
$ABCD$, we are performing inverse processes of the Iwasawa
decomposition.

It should be noted that the decomposition we used in this paper has a
specific purpose. If purposes are different, different forms
of decomposition may be employed.  For instance, decomposition of the
$ABCD$ matrix into shear, squeeze, and rotation matrix could serve
useful purposes for canonical operator
representations~\cite{simon00,naza82}.  The amount of calculation
seems to depend on the choice of decomposition.

Group theory in the past was understood as an abstract mathematics.
In this paper, we have seen that it can be used as a calculational
tool.  We have also noted that there is a place in computer science
for group theoretical tools.

\newpage

\begin{appendix}

\section{Bargmann Decomposition}
In his 1947 paper~\cite{barg47}, Bargmann considered
\begin{equation}\label{alphabe}
W = \pmatrix{\alpha & \beta \cr \beta^{*} & \alpha^{*}} ,
\end{equation}
with $\alpha\alpha^{*} - \beta\beta^{*} = 1$.  There are three
independent parameters.
Bargmann then
observed that $\alpha$ and $\beta$ can be written as
\begin{equation}
\alpha = (\cosh\eta) e^{-i(\phi + \lambda)}, \qquad
\beta = (\sinh\eta) e^{-i(\phi - \lambda)} .
\end{equation}
Then $W$ can be decomposed into
\begin{equation}
W =\pmatrix{e^{-i\phi} & 0 \cr 0 & e^{i\phi}}
\pmatrix{\cosh\eta & \sinh\eta \cr \sinh\eta & \cosh\eta}
\pmatrix{e^{-i\lambda} & 0 \cr 0 & e^{i\lambda}}  .
\end{equation}
In order to transform the above expression into the decomposition
of Eq.(\ref{sp22}), we take the conjugate of each of the matrices with
\begin{equation}
C_{1} =  {1 \over \sqrt{2}} \pmatrix{1 & i \cr i & 1} .
\end{equation}
Then $C_{1} W C_{1}^{-1}$ leads to
\begin{equation}
\pmatrix{\cos\phi & -\sin\phi \cr \sin\phi & \cos\phi}
\pmatrix{\cosh\eta & \sinh\eta \cr \sinh\eta & \cosh\eta}
\pmatrix{\cos\lambda & -\sin\lambda \cr \sin\lambda & \cos\lambda} .
\end{equation}
We can then take another conjugate with
\begin{equation}
C_{2} =  {1 \over \sqrt{2}} \pmatrix{1 & 1 \cr -1 & 1} .
\end{equation}
Then the conjugate $C_{2} C_{1} W C_{1}^{-1} C_{2}^{-1} $ becomes
\begin{equation}
\pmatrix{\cos\phi & -\sin\phi \cr \sin\phi & \cos\phi}
\pmatrix{e^{\eta} &  0 \cr 0 & e^{-\eta}}
\pmatrix{\cos\lambda & -\sin\lambda \cr \sin\lambda & \cos\lambda} .
\end{equation}
This expression is the same as the decomposition given in Eq.(\ref{sp22}).

The combined effect of $C_{2}C_{1}$ is
\begin{equation}
C_{2}C_{1} = {1 \over \sqrt{2}} \pmatrix{e^{i\pi/4} &  e^{i\pi/4} \cr
-e^{-i\pi/4} &   e^{-i\pi/4}} .
\end{equation}
If we take the conjugate of the matrix $W$ of Eq.(\ref{alphabe}) using the
above matrix, the elements of the $ABCD$ matrix become
\begin{eqnarray}
&{}& A = \alpha + \alpha^{*} + \beta + \beta^{*} , \nonumber \\[1.0ex]
&{}& B = -i(\alpha - \alpha^{*} + \beta - \beta^{*}) , \nonumber \\[1.0ex]
&{}& C = -i(\alpha - \alpha^{*} - \beta + \beta^{*}) , \nonumber \\[1.0ex]
&{}& D = \alpha + \alpha^{*} - \beta - \beta^{*} .
\end{eqnarray}
It is from this expression that all the elements in the $ABCD$ matrix
are real numbers.  Indeed, the representation $\alpha\beta$
is equivalent to the $ABCD$ representation.  In terms of the parameters
$\lambda, \eta$ and $\phi$,

\begin{eqnarray}
&{}& A = (\cosh\eta)\cos(\phi + \lambda) +
         (\sinh\eta)\cos(\phi - \lambda)  , \nonumber \\[1.0ex]
&{}& B = (\cosh\eta)\sin(\phi + \lambda) +
         (\sinh\eta)\sin(\phi - \lambda)  , \nonumber \\[1.0ex]
&{}& C = (\cosh\eta)\sin(\phi + \lambda) -
         (\sinh\eta)\sin(\phi - \lambda)  , \nonumber \\[1.0ex]
&{}& D = (\cosh\eta)\cos(\phi + \lambda) -
         (\sinh\eta)\cos(\phi - \lambda)  .
\end{eqnarray}

\end{appendix}

\end{document}